\documentclass[aps,amsfonts,amsmath,prd,preprint,nofootinbib]{revtex4}
\usepackage{epsf,mathrsfs}

\newcommand{\beq}{\begin{equation}}
\newcommand{\eeq}{\end{equation}}

\begin{document}

\title{Holographic multiverse and conformal invariance}

\author{Jaume Garriga$^1$ and Alexander Vilenkin $^2$} \address{ $^1$
  Departament de F{\'\i}sica Fonamental i \\Institut de Ci{\`e}ncies
  del Cosmos, Universitat de Barcelona,\\ Mart{\'\i}\ i Franqu{\`e}s
  1, 08193 Barcelona, Spain\\ $^2$ Institute of Cosmology, Department
  of Physics and Astronomy,\\ Tufts University, Medford, MA 02155,
  USA}

\begin{abstract}

We consider a holographic description of the inflationary multiverse,
according to which the wave function of the universe is interpreted as
the generating functional for a lower dimensional Euclidean theory. We analyze 
a simple model where transitions between inflationary vacua
occur through bubble nucleation, and the inflating part of spacetime
consists of de Sitter regions separated by thin bubble walls. In this model, 
we present some evidence that the dual theory is conformally invariant in the UV.

\end{abstract}

\maketitle

\section{Introduction}

Defining the probability measure in an eternally inflating universe is
one of the key unresolved problems of inflationary cosmology.  Eternal
inflation produces an infinite number of ``pocket universes'', in
which all possible events happen -- not once, but an infinite
number of times.  We have to learn how to regulate and compare these
infinities, since otherwise we cannot distinguish between probable and
highly improbable events, and thus cannot make any predictions at all.

In ``multiverse'' models with a multitude of different vacua, eternal
inflation gives rise to a fractal pattern, where pockets of all possible
vacua are nested within one another.  An important problem in this type of
model is to find the probability distribution for the values of
low-energy constants of nature, such as the parameters of the standard
model of particle physics. Once again, we need some way of regulating
infinities, and the problem here is exacerbated by the complicated
spacetime structure of the multiverse.

The essence of the problem is that the numbers of all kinds of events
in an eternally inflating universe are growing exponentially with
time.  Whatever cutoff method is used, most of the events occur near
the cutoff, and the resulting probability measure depends sensitively
on the cutoff prescription.  This is the so-called ``measure problem''
of inflationary cosmology. (For a review see
\cite{Guth07,Winitzki08}.)

So far, most of the work on the measure problem has been
phenomenological.  Different measure proposals have been examined to
check whether or not they lead to inconsistencies or to a glaring
conflict with the data.  (For recent discussion and references,
see, e.g., \cite{youngness2,LVW,DGSV,DeSimone:2008if}
.)  Other selection criteria have also
been introduced.  For example, it has been argued that the probability
measure should not depend on the initial conditions at the beginning
of inflation.  It seems rather unlikely, however, that this kind of
analysis will lead to a unique prescription for the measure.

A more satisfactory approach would be to motivate the choice of
measure from some fundamental theory.  Attempts in this direction have
been made in \cite{FSY,hat2,Bousso06,GV08,Bousso09}.  In particular, in
Ref.~\cite{GV08} we suggested that the dynamics of the inflationary
multiverse could have a dual description in the form of a
lower-dimensional Euclidean field theory defined at the future
infinity.  The measure of the multiverse can then be defined by
imposing a Wilsonian ultraviolet cutoff $\xi$ in that theory.  We
argued that in the limit of $\xi\to 0$, the boundary theory becomes
conformally invariant, approaching a UV fixed point.  We also argued
that on super-horizon scales the UV cutoff $\xi$ corresponds to a
scale factor cutoff in the bulk theory.

In the present paper we shall further explore the 
holographic duality proposed in Ref.~\cite{GV08}, filling in
some of the missing details.
In Section II we consider the simple model where transitions between different vacua
occur through bubble nucleation, and
the inflating part of spacetime consists of de Sitter regions
separated by thin bubble walls.  We discuss the structure of
future infinity in this model, focusing in particular on the eternal
set ${\cal E}$, defined by eternal timelike curves which always remain
within the inflating region and never encounter terminal bubbles of
negative or zero vacuum energy density.  We show that (i) the
boundary metric on ${\cal E}$ can be chosen to be flat and (ii) the
bubble distribution on the boundary is then approximately invariant
under the Euclidean conformal group, with the invariance becoming
exact in the UV limit.  This supports the conjecture that the dual
boundary theory should be conformally invariant in the UV.

By analogy with AdS/CFT correspondence \cite{Maldacena98,Gubser:98,Witten98}, we
suggested in Ref.~\cite{GV08} that the correspondence between the
multiverse theory in the bulk and its dual on the future boundary is
expressed by the relation
\begin{equation}
\Psi[\bar\phi({\bf x})]\equiv \int D\phi\ e^{iS[\phi]} =
e^{iW[{\bar\phi}]}.  
\label{partition}
\end{equation}
Here, $S$ is the bulk action and the integral is over bulk fields
$\phi$ approaching the prescribed $\phi={\bar\phi}({\bf x})$ at the
boundary.  The amplitude $\Psi[\bar\phi({\bf x})]$ has the meaning of
the wave function of the universe, and $W[{\bar\phi}]$ is the
effective action for the boundary theory with the appropriate
couplings to the external sources ${\bar\phi}$. In Section III we
describe this proposal in more detail, with emphasis on the IR/UV
connection and the implications for the measure problem.

To gain further insight into the properties of the boundary
theory, in Section IV we calculate $W$ for linearized perturbations
around the model of nested bubbles.  First we consider tensor modes in
de Sitter space. This can be used to find correlators like $\langle
\bar h({\bf x})\bar h({\bf x'})\rangle$ when the points ${\bf
x}$,${\bf x'}$ are within the same bubble. The functional form of the
corresponding $W$ is consistent with that expected in a conformal
field theory. Then, we consider fluctuations of the bubble walls in
the approximation where the self-gravity of the bubble can be
neglected. In the boundary theory, the bubble walls mark the
boundaries between regions with different central charge, or different
number of field degrees of freedom. Such boundaries give a
contribution to the trace anomaly which depends on their shape.
We evaluate this contribution from the asymptotic form of the bulk
wave function for the bubble wall fluctuations.  Once again, the
result is consistent with conformal invariance of the boundary theory.

Finally, in Section V we summarize our conclusions and discuss some
open issues.

When this paper was nearly completed, we learned of the work in
progress by Stephen Shenker, Douglas Stanford and Leonard Susskind,
which has some overlap with the ideas presented here.

A related interesting development is the work of Freivogel and Kleban 
\cite{FK09}. They consider a model where bubbles nucleate in a de Sitter
background, and compute correlators for operators which characterize 
the bubble distribution at the future boundary. They find that these
correlators are conformally invariant, and they also discuss a dual 
CFT interpretation of their results.

\section{The model of nested de Sitter bubbles}

\subsection{Flat foliations}

\begin{figure}[ht]

\begin{center}\leavevmode
\epsfxsize=16.5 cm
\epsfbox{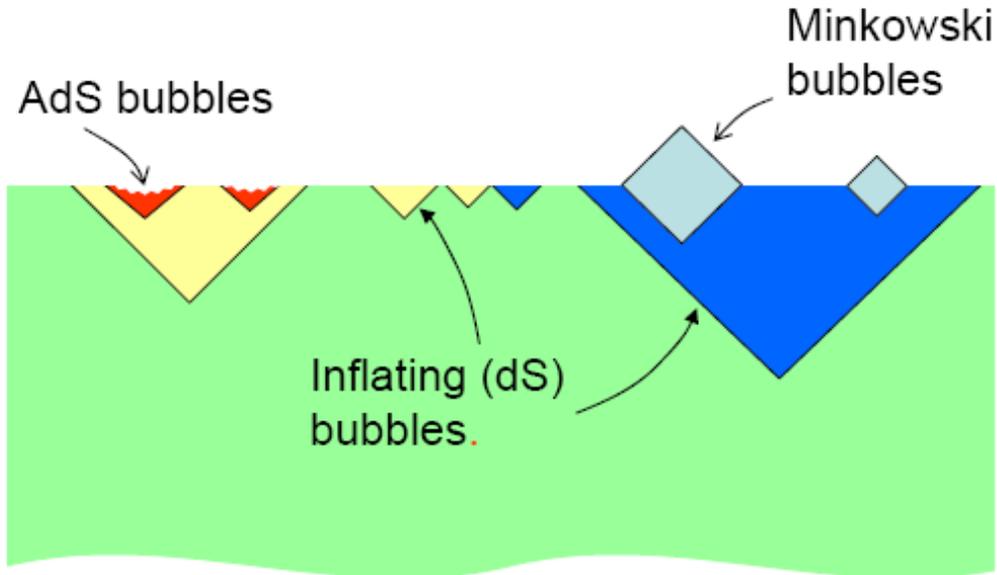}
\end{center}

\caption{Causal diagram of the inflationary multiverse.}
\label{Fig1}
\end{figure}

A diagram illustrating the causal structure of an eternally inflating
spacetime is shown in Fig.~\ref{Fig1}.  Bubbles of all possible types nucleate
and expand, rapidly approaching the speed of light.  The worldsheets
of the bubble walls can therefore be approximated as light cones in
the diagram.  The future boundary of this spacetime includes the
singular boundary corresponding to the big crunch singularities of the
negative-energy anti-de Sitter (AdS) bubbles, ``hats'' corresponding
to the future null and timelike infinities of the Minkowski bubbles,
and the eternal set ${\cal E}$ which is the focus of our interest in
this paper.  AdS and Minkowski bubbles are called ``terminal
bubbles'', since inflation completely terminates in their interiors.
The eternal set ${\cal E}$ consists of the spacelike future boundaries
of the inflating de Sitter (dS) bubbles -- or rather what remains of
these boundaries after we remove the regions eaten up by terminal
bubbles. We can think of ${\cal E}$ as the set of ``endpoints'' of
eternal timelike curves, which never encounter terminal bubbles.  (A
more formal definition will be given in
Section III.)

We shall start by considering a simplified model in which the
inflating part of spacetime consists of pure dS regions separated by
thin bubble walls.  This part of spacetime can be foliated by
spacelike surfaces $\Sigma_t$, labeled by some coordinate $t$.  
The key observation now is that in our simple model of nested dS
bubbles, the spacetime region of interest can be foliated by flat
Euclidean surfaces.  This is not difficult to understand
\cite{recycling}.  A dS space inflating at a rate $H$ can be 
represented as a hyperboloid 
\beq
{\bf X}^2 + Y^2 - T^2 = H^{-2}
\eeq
embedded in a $5D$ Minkowski space.  The flat foliation of this space
is obtained by slicing it with null hyperplanes
\beq
Y-T=const.
\label{Y+T}
\eeq
The resulting metric can be written as.
\beq
ds^2 = - H^{-2}dt^2 + e^{2t} d{\bf x}^2,
\label{flatdS}
\eeq
where $t$ is the scale factor time, that is, 
time measured in units of $H^{-1}$.
(This coordinate system covers only half of the full dS space, but this
is sufficient to cover the interior of any bubble.)  

\begin{figure}[ht]

\begin{center}\leavevmode
\epsfxsize=16.5 cm
\epsfbox{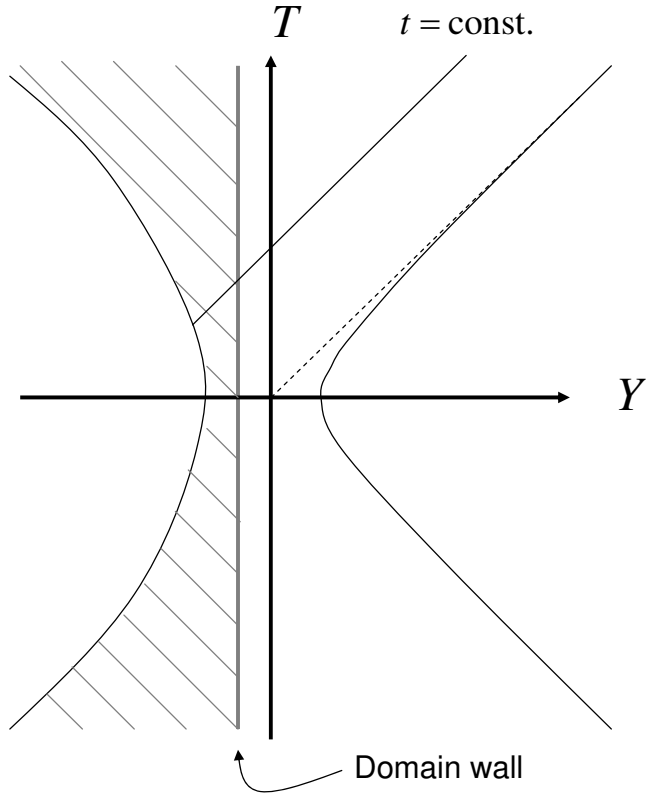}
\end{center}

\caption{The spacetime representing a bubble of lower-energy vacuum
  (shaded) expanding in a higher-energy de Sitter space.  One of the
  flat foliation surfaces is also shown.}
\label{Fig2}
\end{figure}

The spacetime geometry for a bubble of vacuum ${\cal V'}$ in a background of
vacuum ${\cal V}$ can be obtained by matching the hyperboloid (\ref{flatdS})
representing the parent vacuum with a similar hyperboloid for the
daughter vacuum,
\beq
{\bf X}^2 + (Y-Y_0)^2 - T^2 = {H'}^{-2},
\eeq
where the displacement $Y_0$ depends on the tension of the bubble
wall.  The two hyperboloids are joined along a (2+1)-dimensional
hyperbolic surface which represents the worldsheet of the wall (see
Fig.~\ref{Fig2}).  

Once again, a flat foliation can be obtained by slicing this hybrid
construction with null hyperplanes (\ref{Y+T}).  Each slice will then
consist of a spherical region of vacuum ${\cal V'}$ embedded in an infinite
region of vacuum ${\cal V}$ (or vice versa).  Since the spatial
geometry of the slices is flat both inside and outside the bubble, the
volume that is removed from vacuum ${\cal V}$ is equal to the volume of
vacuum ${\cal V'}$ which replaces it.  As a result, the two regions match
without any discontinuity, even though there is a $\delta$-function
curvature singularity along the domain wall in $4D$.

The metric in the bubble interior can also be brought to the form
(\ref{flatdS}), but since the expansion rates in the two vacua are
different, the corresponding time variables do not match at the bubble
wall. In a flat slicing, such as (\ref{flatdS}), the physical radius
of the bubble  
is given by \cite{recycling}
\beq
R^2(t)=(H^{-1}e^t-D)^2+R_0^2, 
\eeq
where $D=(H^{-2}-R_0^2)^{1/2}$ and $R_0$ is the constant 
intrinsic curvature radius of the worldsheet (this constant is determined by
the bubble wall tension and the vacuum energies on both sides of the wall). The radius 
is the same when viewed from inside and from outside, and thus we must have
\beq
t' = t + \ln (H'/H) + O(e^{-t}).
\label{t2t1}
\eeq 
For $t,t' \gg 1$, the time shifts in $t$ and $t'$ are related by
$\Delta t'=\Delta t$ and correspond to equal multiplicative changes
in the scale factors $a(t)=e^t$.  

This argument can be extended to any number of bubbles of any kind
nucleating in the parent vacuum, as well as to bubbles nucleating
within those bubbles, etc.  A slice $\Sigma_t$ through the inflating
region of the multiverse can then be chosen as a $3D$ surface having a
flat Euclidean geometry.  (The label $t$ here can be chosen as the
time variable in the parent vacuum.)  Each bubble which is crossed by
this surface is represented by a spherical region, with bubbles
nucleating later having smaller images on $\Sigma_t$.  The image of an
inflating bubble will generally have images of daughter bubbles nested
within it.  The image will therefore look like a ``sponge'', whose
``pores'' contain vacua of different types.  Some of the ``pores'' may
correspond to other inflating vacua, in which case they will
themselves look like sponges, and so forth.  The ``pores''
corresponding to terminal vacua are represented by ``holes'' in
$\Sigma_t$.  As we move to later and later times $t$, we will see more
and more of this structure.  In the limit of $t\to\infty$, each
``sponge'' becomes a fractal set, with almost all volume eaten up by
terminal bubbles.

We note that the flat foliation we have just described is not the
  same as the scale factor foliation (with the usual definition of the
  scale factor through the expansion along a geodesic congruence).
  The reason is that geodesics undergo some focusing at domain walls
  and the congruence takes some time to adjust to the new expansion
  rate as it crosses from one vacuum to another.  As a result, the
  flat slices deviate from constant scale factor surfaces, typically
  by $\delta t \sim 1$.  This deviation, however, is rather
  insignificant, and we shall refer to the variable $t$
  as the scale factor time.
  
The size distribution for bubbles of different kinds on the flat
slices can be found using the formalism developed in
Refs.~\cite{recycling,GSVW}.  Each bubble can be labeled by two
indices $\{ij\}$, where $i$ refers to the vacuum inside the bubble and
$j$ to the parent vacuum in which it nucleated.  The number of
type-$ij$ bubbles nucleated in a unit comoving volume, $V(t)=e^{3t}$,
in an infinitesimal time interval $dt$ is given by
\beq
dN_{ij}= e^{3t}\lambda_{ij}f_j H_j^{-1}dt.
\label{dNt}
\eeq
Here, $\lambda_{ij}$ is the bubble nucleation rate per unit physical
spacetime volume, $f_j(t)$ is the fraction of space in the $3D$ slice
occupied by vacuum $j$, and there is no summation over $j$.  The
evolution equation for $f_i(t)$ can be written as
\beq
{df_i\over{dt}}=\sum_j M_{ij}f_j,
\label{mastereq}
\eeq
where 
\beq
M_{ij}=\kappa_{ij}-\delta_{ij}\sum_m\kappa_{mi}
\eeq
and
\beq
\kappa_{ij}\equiv\lambda_{ij}{4\pi\over{3}}H_j^{-4}
\eeq
is the probability for a bubble to nucleate per Hubble volume per
Hubble time.

The asymptotic solution of (\ref{mastereq}) at large $t$ has the form
\beq
f_j(t)=f_j^{(0)}+s_je^{-qt}+...
\label{fj}
\eeq
Here, $f_j^{(0)}$ has nonzero components only in terminal vacua, $-q<0$
is the dominant eigenvalue of the matrix $M$ (that is, the largest
nonzero eigenvalue) and $s_j$ is the corresponding
eigenvector.\footnote{It has been shown in \cite{GSVW} that for an
  irreducible landscape of vacua, in which any vacuum can be reached
  from any other nonterminal vacuum by a sequence of transitions, the
  dominant eigenvalue is real, negative, and nondegenerate.}
Substituting (\ref{fj}) into (\ref{dNt}), we have
\beq
dN_{ij}= \lambda_{ij}s_je^{(3-q)t}H_j^{-1} dt.
\label{dNt2}
\eeq

Now, consider a bubble of type $ij$ which was formed at time $t$.  The
radius of this bubble at $t'\gg t$ is
\beq
R(t',t) \approx H_j^{-1}e^{t'-t}.
\label{R}
\eeq
Expressing $t$ in terms of $R$ and substituting in (\ref{dNt2}), we
obtain the bubble distribution
\beq
dN_{ij} = C_{ij}R^{-(4-q)}dR ,
\label{dNR}
\eeq
where $C_{ij}$ is a constant ($R$-independent) coefficient.
We thus see that the size distribution for all bubble types follows the same
power law (\ref{dNR}).

\subsection{Conformal invariance}

It follows from Eq.~(\ref{R}) that a forward time translation by
$\Delta t$ results in a rescaling of (large) bubble sizes by the same
factor $\exp(\Delta t)$.  Since the size distribution of bubbles
(\ref{dNR}) is a power law, the form of the distribution is unchanged
under the rescaling.  

Factoring out the scale factor $e^{t'}$, we can introduce the comoving
bubble radius, 
\beq
r(t',t)=e^{-t'}R(t',t) , 
\eeq
which approaches a constant value at $t'\to \infty$,
\beq
r \to H_j^{-1}e^{-t},
\eeq
where $t$ is the time of bubble nucleation.  In terms of the comoving
coordinates ${\bf x}$, we thus have an
asymptotically static distribution of bubbles at future infinity,
\beq
dN_{ij}\propto r^{-(4-q)}dr.
\label{dNr}
\eeq
This distribution was derived using the asymptotic solution (\ref{fj})
for $f_j(t)$ at large $t$, so we expect it to apply only
approximately, becoming exact in the limit $r\to 0$.  The shape of the
distribution at large $r$ is not universal and is influenced by the
initial conditions at the onset of inflation (e.g., the kind of parent
vacuum we start with, etc.) 

The distribution (\ref{dNr}) is in agreement with scale invariance: in
a space of dimension $d$, a scale invariant distribution has the form
$dN \propto r^{-(d+1)} dr$, while the fractal dimension of the eternal
set ${\cal E}$ is given by $d=3-q$ \cite{GSVW,recycling}.

Another asymptotic symmetry of the bubble distribution is related to
the possibility of choosing different flat foliations of spacetime.
These foliations are related by Lorentz transformations in the $5D$
embedding space, or equivalently, by de Sitter group transformations in the
parent vacuum region.  Each foliation defines a coordinate system
(\ref{flatdS}) and a congruence of timelike geodesics ${\bf x}=const$
in the parent vacuum.  Any two such congruences become asymptotically
comoving (in the region where they overlap) and thus define a
coordinate transformation ${\bf x}\to{\bf x'}$ at the future infinity.  
In the Appendix we show that it is a transformation of the form
\begin{equation}
{x'^i\over {\bf x'^2}} = {x^i\over {\bf x^2}} - {b^i}, 
\label{sct0}
\end{equation}
accompanied by a rotation. (Here, $b^i$ is the dS boost parameter.)
The transformation (\ref{sct0}) is the so-called special
conformal transformation (SCT).  It can be described as a translation
preceeded and followed by inversions.  An important property of SCTs
is that they map spheres into spheres, and thus the spherical shape of
the bubbles is preserved.

By the same argument, a change of foliation induces SCTs in the
asymptotic future of all dS bubble interiors. By continuity accross
bubble boundaries, 
it follows that all transformations should be the
same, so that the entire future boundary ${\cal E}$ is transformed by
a single SCT (plus rotation).  

For any flat slicing, the evolution of the bubble distribution is
described by the same equations
(\ref{dNt}),(\ref{mastereq}),(\ref{R}), although the initial
conditions are generally different.  However, since the late-time
behavior is universal and independent of the initial state, the form
of the distribution should be the same at $t\to\infty$.  This means
that the size distribution of the bubbles should be invariant under
SCTs in the limit of $r\to 0$.

Together with dilatations, translations, and rigid rotations, SCTs
comprise the Euclidean conformal group.  Our conclusion is thus that
the bubble distribution is conformally invariant at $r\to 0$.  If
indeed the multiverse has a dual description on ${\cal E}$, this
suggests that the boundary theory should be conformally invariant in
the UV.

\subsection{Bubble collisions and the persistence of memory}

So far in this discussion we have ignored bubble collisions. These
are interesting in that they are sensitive to initial conditions, an
effect which has been dubbed "the persistence of memory"
\cite{GGV}. This effect can be described as follows. Suppose we have
some initial hypersurface of vacuum of type $i$ which has no bubbles
of any other phase. Then, memory of such initial condition persists
arbitrarily far into the future. Indeed, the preferred congruence
orthogonal to the initial "no bubble" hypersurface defines a rest
frame. An observer in vacuum $i$ at rest in this frame will be equally
likely to be hit by a bubble of type $j$ from any direction in
space. In other words, the probability of collision with new bubbles
is isotropic for this observer. However, if the observer moves with
respect to the preferred congruence, then the observer is more likely
to be hit head on by new bubbles.

When we look at the shapes drawn by the nucleated bubbles at the
future boundary, this translates into the following effect. Let us
first choose our foliation to be (locally) orthogonal to
the preferred congruence.  In these co-moving coordinates, any bubble
which nucleates in the original vacuum $i$ will be equally likely to
be hit from any direction. At the future boundary, this will be seen
as a big spherical bubble "decorated" isotropically by a froth
of smaller bubbles.

However, if the foliation is boosted with respect to the preferred
congruence, then the pattern of bubbles at the future boundary will be
distorted by the corresponding SCT.  Although the spherical shape of a
bubble is preserved by inversions around an arbitrary point, the
isotropic distribution of smaller spheres decorating it will be
shifted preferentially in the direction determined by $b^i$. As a
result, the froth decorating the bubbles will be anisotropic.

Nonetheless, for a given value of the boost parameter $b^i$, the
effect will get smaller for bubbles nucleating at later times.  The
reason is that a congruence orthogonal to any flat foliation becomes
asymptotically comoving to the preferred congruence.  In this way, the
conformal invariance is recovered in the UV.

\subsection{More general foliations}

Flat foliations of the kind we discussed so far are possible in the
simple model of nested bubbles, but cannot be constructed in a general
spacetime.  To get an idea of what the general situation may be like,
we shall now examine a wider class of surfaces.

For a general spacelike surface $\Sigma$, we can introduce a
coordinate system (\ref{flatdS}) in each dS region that this surface
intersects.  In terms of these coordinates, the surface can be represented as
\beq
t = f({\bf x}).
\label{t}
\eeq 
The function $f({\bf x})$ may change its form from one bubble to
another, but we shall require that $\Sigma$ remains smooth at the
bubble walls.  We shall also assume that $f({\bf x})$ is a slowly
varying function, so that
\beq
e^{-f({\bf x})}|\nabla f({\bf x})|\ll H({\bf x}).
\label{Hf}
\eeq
This guarantees that the typical curvature radii of $\Sigma$ are much
greater than the local dS horizon.  In other words, the surface is nearly
flat on the horizon scale.  

The induced metric on $\Sigma$ is 
\beq
ds^2= e^{2f({\bf x})}d{\bf x}^2 -H^{-2}\partial_i f \partial_j f
dx^i dx^j.
\label{induced}
\eeq
The condition (\ref{Hf}) allows us to drop the second term, so we can
write
\beq
ds^2 \approx  e^{2f({\bf x})}d{\bf x}^2
\label{conflat}
\eeq
We thus see that the metric on $\Sigma$ is related to the flat
Euclidean metric by a Weyl rescaling \cite{GV08}.

In a CFT, the observables have well known transformation properties
under Weyl rescalings. Once we know the observables for a given
metric, we can find their values for metrics in the same conformal
class.  In this sense, the boundary theory is Weyl covariant, and the
conformal factor is a gauge redundancy.  We fix the gauge by choosing
a particular metric in the conformal class. For instance, the choice
of a flat boundary metric has the advantage that the conformal
invariance of the theory is manifest in some observables.

\subsection{More general spacetimes}

To get an idea of what the situation is in more general spacetimes, not necessarily piecewise
de Sitter, let us consider the case of a metric which is locally FRW, but with the rate of 
expansion varying from place to place on scales much bigger than the horizon. To be more precise,
let us consider the metric of a spacetime whose expansion is isotropic. This means that the 
extrinsic curvature of equal time slices orthogonal to a geodesic congruence is proportional to 
the spatial metric on the slices. The metric can then be written as
\begin{equation}
ds^2 = -d\tau^2 + e^{2 N} \gamma_{ij}({\bf x}) dx^i dx^j,
\end{equation}
where $N(\tau,{\bf x})$ can be interpreted as the number of e-foldings in proper time gauge.
Introducing $\eta = - e^{-N}$, we have
$$
d\eta = -\eta dN = -\eta (H d\tau + N,_i dx^i),
$$
where $H=N,_\tau$. Hence,
\begin{equation} 
ds^2 = \eta^{-2} [-H^{-2}( d\eta + \eta N,_i dx^i)^2 +  \gamma_{ij} dx^i dx^j].
\end{equation}
In the limit $\eta\to 0$, the second term within the round brackets can be neglected,\footnote{In neglecting this
term, we are implicitly assuming that the metric is smoothed over a fixed co-moving scale. The motivation for
this smoothing will be clear from the discussion in the next Section, since the UV cut-off of the boundary theory
corresponds to fixed co-moving scale.} and we
have 
\begin{equation}
ds^2 \approx \eta^{-2} [-H^{-2} d\eta^2 + \gamma_{ij} dx^i dx^j]. 
\end{equation}
Thus, we find that near the future boundary 
the metric of the inflating part of space-time is conformal to 
\begin{equation}
d\hat s^2 = -H^{-2}(\eta,{\bf x}) d\eta^2 + \gamma_{ij}({\bf x}) dx^i dx^j.\label{conf}
\end{equation}
This suggests that we can identify ${\cal E}$ with the future conformal boundary
of the metric (\ref{conf}). Clearly, $\gamma_{ij}$ plays the role of the metric at the conformal boundary.
The arbitrariness in the choice of the congruence translates into Weyl rescalings of the metric at the 
conformal boundary. Indeed, in the inflating background, any two geodesic congruences become asymptotically 
co-moving to 
each other. Consider two geodesics that have the same endpoint ${\bf x}$ at the conformal boundary. 
By the time their relative speed becomes completely negligible, the number of e-foldings from the initial 
surface for both geodesics will be different by some $\Delta N({\bf x})$, so in the asymptotic future we have 
$\eta=\eta' e^{2\Delta N({\bf x})}$. Hence, the spatial metric on surfaces of very small constant $\eta$
will be related to that on surfaces of very small constant $\eta'$ by Weyl rescaling.

\section{The wave function of the universe and the IR/UV 
correspondence}

The wave function of the universe can be expressed as a path integral
\beq 
\Psi[{\bar\phi}]=\int_{\Sigma_i}^{\Sigma} D\phi e^{iS}, 
\label{wavepath}
\eeq 
where the integration is over the bulk fields $\phi(x)$ interpolating
between some initial conditions on hypersurface $\Sigma_i$ and
approaching the prescribed values ${\bar\phi}({\bf x})$ on the surface
$\Sigma$, which plays the role of the future boundary of the spacetime  
region of integration.  The nature of the initial conditions on
$\Sigma_i$ is the subject of some debate (for a review see \cite{discord}
and references therein), but it will not be important for our
discussion here.  The reason is that an eternally inflating region of
spacetime quickly forgets its initial conditions.  The memory of the
initial state is retained only on the largest comoving scales, while
on smaller scales the evolution exhibits an attractor behavior.  For
example, we saw in Section II.A that the asymptotic bubble
distribution (\ref{dNr}) at short distances is independent of the initial conditions.
This distribution is reached in any comoving region including at least
one eternal geodesic.

Thus, to study the asymptotic small-scale behavior, we do not need to
invoke the wave function for the entire universe.  We could start,
say, with a cubic region filled with vacuum $i$ and having size of a
few $H_i^{-1}$.  The conditions at the boundaries of this region are
not important; we could, for example, impose periodic boundary
conditions.  
In our simple model, the spacetime consists of nested regions of dS
space separated by thin walls, and the evolution is nearly classical,
except for occasional nucleation of bubbles.  According to the
discussion in Section II.A, this spacetime can be foliated by flat
hypersurfaces, so the initial and final surfaces can be chosen to be
flat.  The future boundary can then be approximated by one of
the surfaces $\Sigma_t$ and the only variables that need to be
specified on that boundary are the types of vacua and the centers and
radii of the corresponding bubbles.  We expect that in the limit
$t\to\infty$ all correlations of physical significance will not depend
on the choice of the initial vacuum on $\Sigma_i$.

The conjectured correspondence (\ref{partition}) relates the wave
function $\Psi[{\bar\phi}]$ for the bulk theory to the effective
action of the boundary theory.  The surface $\Sigma$, with regions
covered by terminal bubbles removed, gives an approximate
representation of the eternal set ${\cal E}$, as we describe below.
As we shall see, the boundary theory, regularized with an
appropriate UV cutoff, can also be thought of as living on that
surface.

\subsection{The discrete boundary theory and its continuum limit}

A point $P\in {\cal E}$ can be thought of as the endpoint of an
eternal timelike curve (that is, a curve that never encounters
terminal bubbles).  More precisely, ``points'' or elements of ${\cal
E}$ are identified with chronological pasts of eternal timelike curves
\cite{HawkingEllis}.  Two curves with the same past define the same element of
${\cal E}$. Physically, an element of ${\cal E}$ is the equivalence
class of eternal time-like curves which remain forever in mutual
causal contact. An eternal curve can be pictured as the worldline of
an observer and the corresponding boundary ``point'' $P$ is the
spacetime region inside the causal horizon of that observer (see
Fig.~{\ref{Fig3}).\footnote{Elements of ${\cal E}$ can also be identified with
points in the spacelike part of the future conformal infinity, like
the point $P$ in the conformal diagram in Fig.~\ref{Fig3}.  This point is the
endpoint of the eternal timelike curve $\gamma$, and the past of
$\gamma$ appears in the diagram as the interior of the past light cone
of $P$.  This definition of ${\cal E}$ is possible only if the
spacetime admits a conformal infinity, which may not be the case in general.}

\begin{figure}[ht]

\begin{center}\leavevmode
\epsfxsize=16.5 cm
\epsfbox{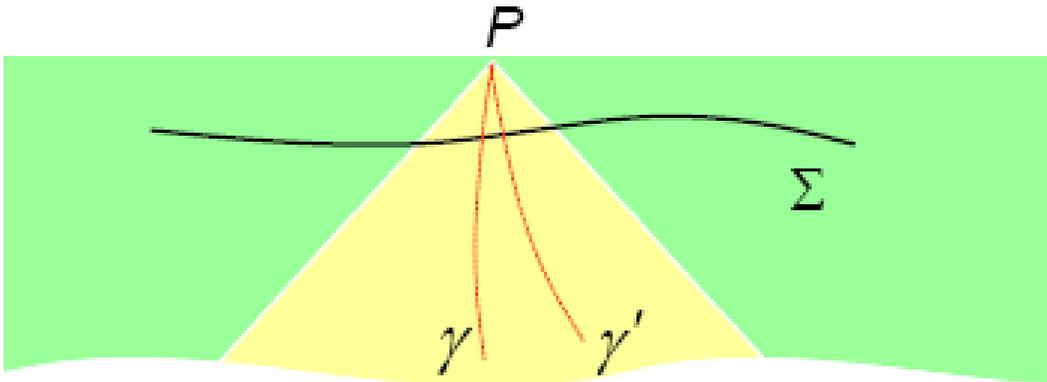}
\end{center}

\caption{Eternal curves $\gamma$ and $\gamma'$ have the same past and
  thus define the same point $P\in {\cal E}$.  The intersection of
  this past (appearing in the figure as the past light cone of $P$)
  with a spacelike surface $\Sigma$ gives the image of $P$ on
  $\Sigma$.} 
\label{Fig3}
\end{figure}

The image $P_\Sigma$ of a point $P\in {\cal E}$ on the surface
$\Sigma$ is given by the intersection of $\Sigma$ with $P$ (see
Fig.~\ref{Fig3}).  In a pure dS space, $P_\Sigma$ would be a spherical region
of radius equal to the dS horizon.  In our model, the shape of
$P_\Sigma$ will be distorted by intervening bubbles, but its size will
still be roughly given by the local horizon.  (If $P_\Sigma$ covers
more than one vacuum, its size is typically set by the largest of the
corresponding horizons.)  In this way the entire eternal set ${\cal
E}$ is mapped onto $\Sigma$, with each point represented by a roughly
horizon-size region.  

Now, instead of considering the boundary theory at ${\cal E}$, let us 
consider a boundary theory defined on $\Sigma$, which should approximate
the theory at ${\cal E}$ on scales larger than a certain Wilsonian cut-off.

First of all, we note that from the point of view of the theory living
on $\Sigma$, the maximum possible resolution is the size of the images
$P_\Sigma$ of the points $P$ of the future boundary. This resolution
is position dependent,
\beq
l_c({\bf x})\sim H^{-1}({\bf x}),
\label{lcH}
\eeq
where $H^{-1}({\bf x})$ is the local dS horizon radius.  It is
interesting to note that, with this choice, the number of degrees of
freedom of the boundary theory on $\Sigma$ corresponds to that of the
bulk theory.\footnote{In the present discussion, and following
standard practice, number of degrees of freedom is synonymous with the
logarithm of the number of quantum states that the system can be in.}
The argument here is similar to that of Susskind and Witten for
AdS/CFT \cite{SW98}.  

The number of bulk degrees of freedom in a
horizon region of vacuum $i$ is given by the Gibbons-Hawking entropy
of the dS space,  
\beq
{\cal N}_{bulk}\sim H_i^{-2}.
\label{Nbulk}
\eeq
To estimate the corresponding number ${\cal N}_{boundary}$ in the
boundary theory, we note that, with the cutoff scale at $l_c = 1/H$,
the energy density of each field at the boundary cannot much exceed
$1/H^4$.  This means that a horizon volume can contain at most $\sim
1$ quanta of frequency $\sim H$.  Longer wavelength modes contribute
much less to the entropy, so we can disregard them.  Thus, each field
carries about one bit of information per horizon, and the entropy per
horizon is of the order of the number of fields.\footnote{We thank
Raphael Bousso for clarifying this point to us.}  For a conformal
field theory, this number is related to the central charge of the
theory.  For the CFT dual to dS space, it has been estimated as
\cite{Strominger1,Strominger2}
\beq
{\cal N}_{boundary}\sim H_i^{-2}.
\eeq
(see also the discussion in Section IV.)
Thus, we see that, with $l_c\sim H_i^{-1}$, we have
\beq
{\cal N}_{bulk} \sim {\cal N}_{boundary}.
\eeq

The theory on $\Sigma$ can be thought of as a field theory only on
scales larger than $l_c$, which plays a role analogous to the lattice
spacing in a discrete system. Borrowing the terminology of
Ref.~\cite{hat2}, we can call this irregular lattice ``a fish
net''.  Each vertex of this fish net carries ${\cal N}_{boundary}$
"internal" degrees of freedom.

The continuum limit on $\Sigma$ is obtained for physical wavelengths
longer than a certain Wilsonian cutoff $\xi_0$, which should be much
bigger than all ``lattice spacings'' $H_i^{-1}$,
\beq
\xi_0 \gg L_0= {\rm max}_i H_i^{-1},
\label{lc}
\eeq
where the maximization is over all dS vacua.  In the nested dS
model, we can choose $\Sigma$ to be one of the surfaces
$\Sigma_t$; then the Wilsonian cutoff $\xi$ in co-moving coordinates
must satisfy
\beq
\xi \gg L(t)\equiv L_0\ e^{-t}.
\label{xi}
\eeq
Here, $L(t)$ can be thought of as a physical UV cutoff, where the
continuum limit breaks down.  This cutoff shrinks as the bulk boundary
surface $\Sigma$ is moved to the future.  As a result, at later times
the eternal fractal ${\cal E}$ can be seen at a greater and greater
resolution.  This situation is analogous to the IR/UV correspondence
in the AdS/CFT \cite{SW98}, with the late-time boundary playing the role
of the IR cutoff in the bulk.  
The scale-factor time evolution in the bulk corresponds to the renormalization group 
flow in the boundary theory.\footnote {For the case of Euclidean
AdS, the Callan-Symanzik RG flow equations for the boundary theory can be derived from a 
semiclassical expansion of the Wheeler-de Witt equation in the bulk \cite{BVV},
with the scale factor playing the role of the renormalization scale. 
Similar considerations were presented in Refs. \cite{VS,LMN} for the case of slow-roll inflation.} 
The theory on ${\cal E}$ is obtained in the limit $t\to
\infty$, where $L(t)\to 0$ and there is no physical UV cutoff.

In our discussion so far we assumed the spacetime to the future of
$\Sigma$ to be well defined and classical.  Of course, this is not so,
even in the nested dS model.  Spontaneous nucleation of bubbles allows
for a multitude of possible histories, and the mapping of ${\cal E}$
onto $\Sigma$ will generally be different for different histories.  We
note, however, that these differences affect the mapping only on
scales smaller than or comparable to the causal horizon, which in turn
is smaller than $L_0$.  So, with the choice of the cutoff length
satisfying (\ref{lc}), the mapping is insensitive to the future
evolution.

The picture suggested by the above discussion is the following. The
hypersurface $\Sigma$ can be thought of as a discrete system
consisting of a juxtaposition of ``lumps" $P_{\Sigma}$.  From a bulk
point of view, a lump is the interior of the causal horizon of an
eternal observer.  This has a finite number of internal degrees of
freedom, proportional to the horizon area (it may therefore be
appropriate to visualize $P_\Sigma$ as a horizon sized closed shell
embedded in $\Sigma$, with the degrees of freedom living on the
shell). At distance scales much larger than the size of the lumps, a
continuum description emerges. The co-moving size of the lumps gets
smaller as the reference surface $\Sigma$ is pushed forward in time,
and in the limit when it is sent to future infinity, the continuum
description becomes valid all the way to the UV (where it should be
conformally invariant).

\subsection{The boundary measure}

In Ref. \cite{GV08} we proposed that a measure for calculating
probabilities in the multiverse may be formulated at future
infinity ${\cal E}$. The idea was to use the Wilsonian cutoff $\xi$
of the boundary theory to render the number of events finite. Any
event occurring in the bulk, which requires for its description a bulk
resolution corresponding to a physical wavelength $\lambda_{min}$,
will leave an imprint 
on ${\cal E}$.  The resolution which is
needed to reconstruct the event from its imprint at the future
boundary will be given by
\begin{equation}
\xi=\lambda_{min}\ e^{-t_*}, 
\label{sfc}
\end{equation}
where $t_*$ is the scale factor time of the event.  For a fixed bulk
resolution $\lambda_{min}$, the Wilsonian UV cutoff $\xi$ of the
boundary theory corresponds to an IR scale factor cutoff at $t=t_*$ in the bulk, 
with $\xi$ and $t_*$ related by Eq.~(\ref{sfc}).
Note that $\lambda_{min}$, which plays the role of a Wilsonian UV cut-off in the bulk, depends in 
principle on what kind of event we are interested in, and the cutoff time $t_*$ will have that dependence 
as well.  An alternative approach is to set
$\lambda_{min}\sim 1$, so that all events above the Planck scale will
be resolved at $t<t_*$.
\footnote{The correspondence between the boundary cut-off and the scale factor cut-off is only approximate. The 
main reason is that the scale factor time is not well defined on subhorizon scales. If one uses the density of a 
dust of test particles in order to define the expansion, then this becomes ill defined in regions of structure formation. 
One can use other definitions, but the choice is not unique. Moreover, Eq. (\ref{sfc}) assumes that wavelengths are 
conformally stretched with the expansion. This is certainly the case on superhorizon scales, but on subhorizon scales 
the wavelengths of signals can be affected by causal processes other than the expansion of the universe.}


Our prescription for the boundary measure is not limited to the
nested dS model or flat foliations.  In the general case, we can use a
scale factor foliation, starting with some surface $\Sigma_0$, which
is smooth on scales larger than $L_0$ and otherwise arbitrary.  The
scale factor $a$ is defined as the expansion factor along the
congruence of geodesics orthogonal to $\Sigma_0$, and the scale factor
time is defined as $t = \ln a$.  As before, each surface of constant
$t$ is the site of a boundary theory with a UV cutoff $\xi$ satisfying
(\ref{xi}), and the scale-factor time evolution corresponds to RG
flow in the boundary theory.

The scale factor time is not a good foliation parameter in regions of
structure formation, where geodesics converge and cross.  However,
these phenomena affect only sub-horizon scales, which are smoothed out
by the super-horizon cutoff $\xi$.  We require, therefore, that the
foliation surfaces $\Sigma_t$ should also be smooth on scales $\lesssim
\xi$.

Different choices of the initial surface $\Sigma_0$ are related
by Weyl transformations on ${\cal E}$.  The freedom of choosing this
surface can be used to obtain a foliation with desired properties.
A flat foliation, like the one we discussed for the nested dS model,
is possible only in very special cases.  Its closest analogue in a
more general spacetime is a foliation by surfaces having a vanishing
Ricci scalar,
\beq
R^{(3)}=0.
\label{R=0}
\eeq
In a spacetime which is locally FLRW with small perturbations, this
condition can always be satisfied by a position-dependent time shift.
If we choose the initial 3-surface $\Sigma_0$ satisfying (\ref{R=0}),
it can be shown that this property is preserved (to linear order) by
scale factor time evolution \cite{Sasaki}.

\section{The wave function and conformal invariance}
 
In this Section we go beyond the simple model of nested bubbles
discussed in Sec.II and allow some perturbations in the inflating
regions and on the spherical bubble walls. First, we consider massless
fields (or metric perturbations) in the case of a single vacuum. Then,
we consider the bubble walls which separate domains with different
vacua. In general, the bubble wall fluctuations are entangled with the
metric perturbations. Here, we shall consider bubble fluctuations in
the region of parameter space where the self-gravity of the bubble can
be ignored and the two types of perturbations can be
disentangled.  As we shall see, the effective action which is
obtained through the correspondence (\ref{partition}) does have some
of the features expected in a CFT.

For the case of massless fields in de Sitter, our discussion closely
follows that of Maldacena in Ref.~\cite{Maldacena02}. A formal
diference is in the expression of the conjectured dS/CFT
correspondence. We have $e^{iW}$ on the right-hand side of
(\ref{partition}), while Ref. \cite{Maldacena02} has $e^{-W}$. If the
effective action is expressed in terms of a boundary path integral, we
propose that this should be written in the form
\begin{equation}
e^{i W[\bar\phi]} = \int_{\cal E} D\psi e^{i\bar S[\psi, \bar\phi]}.
\end{equation}
There is no time variable at the boundary, and in this sense the
action $\bar S$ for the boundary fields $\psi$ is Euclidean. However,
instead of using $e^{-\bar S}$, as in standard Euclidean theories
which are obtained by Wick rotation from Lorentzian time, here we
propose using the phases $e^{i\bar S}$. This is because the wave
function $\Psi$, given by the bulk path integral in (\ref{partition}),
is complex.\footnote{This is in contrast with the 
Hartle-Hawking wave function, which is real.}  In fact, the phase of
the wave function grows at late times in proportion to powers of the
scale factor. In the boundary theory, these growing phases can be
interpreted as ultraviolet divergences, which can be removed by local
counterterms in $\bar S$.  With our conventions, the counterterms are
real.

\subsection{Massless fields in de Sitter}

Let us consider the wave function of a free massless scalar field $h$
in de Sitter.  The same wave function describes linearized gravitons
in the transverse traceless gauge.  For the sake of comparison with
existing CFT calculations, let us work with a (d+1)-dimensional de
Sitter space. In flat conformal coordinates, the metric is given by
\begin{equation}
ds^2= a^2(\eta) [-d\eta^2+ d{\bf x}^2],
\end{equation}
with $a(\eta)=-1/H\eta$.
Decomposing the field in Fourier components
\begin{equation}
h({\bf x}) = \int d^d{\bf k} {e^{i{\bf k}{\bf x}}\over (2\pi)^{d/2}}
h_{\bf k},
\end{equation}
the Gaussian solution of the Schrodinger equation takes the form $\Psi
= e^{iW}$, with
\begin{equation}
W= \int d^d{\bf k}\ \left({a^{d-1}\over 2}\ {v'_{\bf k}\over v_{\bf
k}}|h_{\bf k}|^2 + i \ln v_{\bf k}\right).\label{gaussian}
\end{equation}
Here a prime indicates derivative with respect to $\eta$, and $v_{\bf
k}(\eta)$ is a solution of the massless wave equation $v''_{\bf
k}+(d-1)(a'/a)v'_{\bf k}+{\bf k}^2 v_{\bf k}=0$, which should be of
``negative frequency", $v^*_{\bf k}v'_{\bf k}-v_{\bf k}v'^*_{\bf k}=i
a^{1-d},$ so that the Gaussian is normalizable.  The Bunch-Davies
vacuum corresponds to the choice
\begin{equation}
v_{\bf k}(\eta)={\pi^{1/2}\over 2H^{1/2}}a^{-d/2}(\eta)
H^{(1)}_{d/2}(k\eta), \label{hankel}
\end{equation}
where $H^{(1)}_{d/2}$ is the Hankel function. 

Let us consider the case $d+1=5$. At late times, after modes have
crossed the horizon $(-k\eta \ll 1)$, Eq. (\ref{gaussian}) takes the
asymptotic form
\begin{equation}
W[\bar h({\bf x})]= {1\over 2}\ \int d^4{\bf k}\left( {-k^2 a^2 \over
2 H} + {k^4\over 8 H^3} [\ln (k^2/H^2 a^2)+i\pi+2\gamma] +
O(a^{-2})\right)\label{g2} |h_{\bf k}|^2 + ...,
\end{equation}
where the ellipsis denote terms which are independent of
$h_{\bf k}$.  The structure of (\ref{g2}) is similar to the one
obtained in the context of AdS/CFT by Gubser, Klebanov and Polyakov
\cite{Gubser:98}, except that, here, $W$ contains an imaginary
part. The imaginary part should be there because the amplitude of
perturbations on superhorizon scales is determined by $|\Psi|^2=e^{-2
Im [W]}$. This has a well defined limit as $a\to \infty$,
\begin{equation}
|\Psi|^2 \propto \exp\left[-\int d^4{\bf k}\left({\pi\over 8
H^3}k^4\right) |h_{\bf k}|^2\right],
\end{equation}
corresponding to a scale-invariant spectrum of perturbations $\langle
h^*_{\bf k}h_{\bf k'}\rangle = (8H^3/\pi k^{4}) \delta({\bf k'}-{\bf
k})$.  Let us now consider the real part of $W$. The first term
diverges like $a^2$ as we approach the future boundary. This term is
analytic in $k^2$, and its coefficient can be changed by adding
boundary counterterms of the form $\int (\partial_i h)^2 d^4 {\bf
x}$. The divergence of the kinetic term as $a\to \infty$
suggests that the field is non-dynamical at the boundary.
Introducing $\mu=-\eta^{-1}=Ha$, the non-local part is written as
\begin{equation}
Re[W]= {H^{-3}\over 16} \int d^4{\bf k}\ k^4 \ln (k^2/\mu^2)\ |h_{\bf
k}|^2 + {\rm analytic}.\label{k4logk}
\end{equation}

As mentioned above, the wave function for linearized gravitons in the
traceless and transverse gauge is exactly the same as that for the
massless scalar field $h$. The expression (\ref{k4logk}) we obtain for
$W$ does indeed take the standard form of an effective action for a
conformal theory coupled to an external gravitational field in 4
dimensions. For the case of free fields, this was first discussed by
Tomboulis \cite{Tomboulis:77}. 

The structure of (\ref{k4logk}) is not difficult to understand.
The renormalized effective action for a CFT propagating in a curved
background takes the form \cite{BD}
\footnote{This can be understood as follows. A change in the 
renormalization scale $\mu \to \tilde \mu$ is equivalent to a global
rescaling of the metric $g_{ij}\to \Omega^2 g_{ij}$, with
$\Omega=\tilde \mu/\mu$. Hence $dW_{ren}/d\ln \mu =  \int (\delta
W_{ren}/\delta\ln\Omega) d^d x$.  Since the trace anomaly
$\langle T\rangle=(1/\sqrt{-g})\delta W_{ren}/\delta\ln\Omega$ should be
independent of $\mu$, it follows that $W_{ren}$ is linear in
$\log\mu$, which leads to the form (\ref{wren}).  We thank Igor
Klebanov for a discussion of this point.}
\begin{equation}
W_{ren} = a_{d/2} \ln\mu^2 + ... \label{wren}
\end{equation}
where the ellipsis indicate terms which are independent of $\mu$, and
\begin{equation}
a_{d/2}=\int \langle T\rangle  \sqrt{g}\ d^d x 
\end{equation}
is the integrated trace anomaly in $d$ dimensions.
When $d$ is even, $a_{d/2}$ is given in
terms of an integral of geometric invariants
constructed from contractions of the Riemann tensor. In
$d=4$, this takes the form
\begin{equation}
a_2 = \int d^4 x \sqrt{g} \left[c_1 R^2 + c_2 R_{ij} R^{ij}+c_3
R_{ijkl} R^{ijkl} \right].
\label{bv2}
\end{equation}
The coefficients $c_i$ are such that $a_2$ is a linear combination of
the Euler number and the integral of the Weyl tensor squared. Both are
invariant under Weyl rescalings (see e.g. \cite{BD} and references
therein). More generally, $a_{d/2}$ is Weyl invariant in $d$
dimensions.\footnote{This follows from (\ref{wren}), by noting that the
functional derivative of $W_{ren}$ with respect to $\ln\Omega$ is the
anomalous trace $\langle T \rangle$, which should not depend on
$\mu$.}  To quadratic order in the metric perturbation $h_{ij}$, this
leads to the structure $c_i\ k^4 |h_{\bf k}|^2 \ln \mu^2$ in momentum
space. The form of the non-analytic piece in (\ref{k4logk}) is
recovered by noting that the argument of the logarithm must be
dimensionless.

The numerical constants $c_i$ in front of the geometric
invariants scale in
proportion to the number of fields (central charge) in the
CFT. Comparing with (\ref{k4logk}) we have
\beq
c \sim H^{-3}.  
\eeq
Parametrically, this is also the entropy of the
bulk de Sitter space
\cite{Strominger1,Strominger2,Maldacena02}.

In the case of a 4 dimensional bulk $(d+1=4)$, substituting
(\ref{hankel}) into (\ref{gaussian}), and expanding at late times
$(-k\eta\ll 1)$, one obtains
\begin{equation}
W[\bar h({\bf x})]= {1\over 2}\ \int d^3{\bf k}\left( {-k^2 a \over H} 
+i {k^3\over H^2} + O(a^{-1})\right)
|h_{\bf k}|^2 + ...,\label{g3}
\end{equation}
The first term, which diverges as we approach the future boundary, is
analytic, while the second one is non-local and imaginary. The second
term leads to the familiar expression for the scale invariant
amplitude of tensor modes $\langle h^*_{\bf k} h_{\bf k'}\rangle = H^2
k^{-3} \delta({\bf k'}-{\bf k})$.  In the present case, the Hankel
functions are of half integer order, and do not lead to logarithmic
terms in the asymptotic future. From the CFT point of view, this is in
agreement with the fact that there is no trace anomaly in odd
dimensions, and so there are no logarithmic divergences. The number of
fields in the CFT can be estimated from the coefficient of the
non-analytic piece in $W$. Again, this scales in proportion to the
entropy of the bulk de Sitter space \cite{Strominger1,Strominger2,Maldacena02},
\beq
c\sim H^{-2} .
\eeq

\subsection{Bubble fluctuations.}

Throughout this subsection we work in the approximation where gravity
of the bubble is unimportant. In the thin wall limit, we require that
\begin{equation}
T R_0 \ll 1, \quad (\Delta\rho_V) R_0^2 \ll 1. \label{instgrav} 
\end{equation}
where $T$ is the tension of the bubble wall, $\Delta\rho_V$ is the
energy density gap between the inside and the outside of the bubble
wall, and $R_0$ is the intrinsic curvature radius of the
worldsheet. The conditions (\ref{instgrav}) mean that the geometry
on the scale of the horizon is not appreciably distorted due to
gravity of the bubble wall and due to the change in the vacuum energy.
Under these approximations, $R_0$ is given by \cite{Berezin:87,rates}
\begin{equation}
R_0^2 \approx {(p+1)^2T^2\over (p+1)^2 H^2 T^2+ (\Delta
\rho_V)^2}. \label{radius}
\end{equation}
For reference, we have written the expression for arbitrary worldsheet
dimension $p+1$, although we are mostly interested in the case of
ordinary membranes $p=2$.  Note that we allow the energy gap
$\Delta\rho_V$ to be significant, even comparable to the energy
density in the false vacuum, as long as (\ref{instgrav}) is satisfied.

The interest of this weak backreaction limit is that we can study
fluctuations of the bubble wall without the need of coupling these to
metric fluctuations. We will call this the ``Goldstone limit", since
the only relevant degree of freedom is the one associated with normal
displacements $\delta x^{\mu}$ of the worldsheet $x^{\mu}(\xi^i)$.  In
the flat chart of de Sitter space,
\begin{equation}
ds^2 = -dt^2 + e^{2Ht} (dr^2 + r^2 d\Omega^2)
\end{equation}
the trajectory of a bubble wall centered at the origin of coordinates
is given by
\begin{equation}
r_w^2(t)=H^{-2} (e^{-2Ht_0} + e^{-2Ht}) - 2 H^{-1}
(H^{-2}-R_0^2)^{1/2} e^{-H(t+t_0)},
\end{equation}
where $t_0$ is a free parameter which can be interpreted as the time
of nucleation. The unit vector $n^{\mu}$ normal to the wall
worldsheet is given by
\begin{eqnarray}
n^t&=& R_0^{-1} \left( (H^{-2}-R_0^2)^{1/2}- H^{-1} e^{-H(t-t_0)}\right),\\
n^r&=& R_0^{-1} r_w(t) e^{-H(t-t_0)}.\label{normals}
\end{eqnarray}
We now parametrize the normal displacement of the wall 
as \cite{GV91}
\begin{equation}
\delta x^{\mu} = T^{-1/2} \phi\ n^{\mu}. \label{displacement}
\end{equation}
The factor $T^{-1/2}$ is introduced so that $\phi$ is
a worldsheet scalar with standard normalization. The linearized action
for this scalar consists of a canonical kinetic term and a tachyonic
mass term:
\begin{equation}
m_\phi^2 = -(p+1) R_0^{-2}.
\end{equation}

Let us now study the wave functional for this scalar field $\phi$. The
worldsheet of the bubble is itself a de Sitter space of curvature
radius $R_0$ and dimension $p+1$. This worldsheet will cut a 
sphere at future infinity, and for that reason it is convenient to
use the closed chart for the $p+1$ dimensional de Sitter spacetime,
with coordinates $\xi^i = (\tilde\eta,\Omega_a)$, with $a=1,...,p$,
and metric given by
\begin{equation}
ds^2_w = \tilde a^2 \left(- d\tilde\eta^2 + d\Omega_{p}^2 \right).
\end{equation}
Here
\begin{equation}
\tilde a = R_0/\cos\tilde \eta,
\end{equation}
and $d\Omega_p$ is the metric on a unit p-sphere.  Expanding the
normal displacement as
\begin{equation}
\phi= \sum_{LM}  \phi_{LM} Y_{LM}(\Omega),
\end{equation}
where $Y_{LM}$ are spherical harmonics, the wave functional is given
by $\Psi=e^{iW}$, with
\begin{equation}
W= \sum_{LM} \left({\tilde a^{d-1}\over 2}\ 
{v'_{L}\over v_{L}}|\phi_{LM}|^2 + i \ln v_{L}\right).\label{gaussian2}
\end{equation}
The Bunch-Davies vacuum in the closed chart corresponds to the choice
\begin{equation}
v_{L} = A_L (\cos \tilde \eta)^{p/2} 
\left(P^\nu_{L-1+p/2}(\sin\tilde \eta)+{2i\over \pi} Q^\nu_{L-1+p/2}
(\sin\tilde \eta)\right),
\label{legendre}
\end{equation}
where $A_L$ is a normalization constant.  The index of the Legendre
functions $P$ and $Q$ is given in terms of the mass of the field as
\begin{equation}
\nu = \left({p^2\over 4} - m^2_{\phi} R_0^2\right)^{1/2} = {p+2 \over 2}
\end{equation}

For bubbles in the 3+1 dimensional multiverse we have $p=2$, and
$\nu=2$.  Inserting (\ref{legendre}) into (\ref{gaussian2}) and
expanding for late times ($\tilde a \to \infty$, or $\tilde \eta\to
\pi/2$), we have \footnote{To simplify this expression, we have used a
momentum independent field redefinition at late times $\phi \equiv
(\sin\tilde \eta)^{1/2} \hat\phi$, when $\sin\tilde \eta = 1+0(\tilde
a^{-2})$. In the expression above we have not written the hat on top
of $\phi_{LM}$, since we will be interested in the behaviour at the
future boundary, where both variables coincide.}
\begin{equation}
W = \sum_{LM} \left( {\tilde a^2 \over 2 R_0} + {R_0\Delta \over 4} +
{R_0^3\Delta (\Delta+2)\over 16\ \tilde a^2}  
\left[\ln \left({R_0^2\over 4\tilde a^2}\right)+2\left(\psi(L)+
{1\over L}\right)+ i\pi+2\gamma\right]
\right)
|\phi_{LM}|^2,\label{g3}
\end{equation}
where we have omitted terms of order $\tilde a^{-4}$ (as well as terms
independent of $\phi$).  Here, $\psi$ is the digamma function, and we
are using $\Delta \equiv -L(L+1)$ to denote the Laplacian
eigenvalues. In the
limit of large $L$,  $\psi(L) \approx  \ln L$, so the non-analytic term in (\ref{g3})
is proportional to  $L^4 \ln (R_0^2 L^2 / 4\tilde a^2)$.
The formal similarity between (\ref{g2}) and (\ref{g3})
is to be expected, since the index $\nu=2$ of the Legendre functions
which characterizes a field with $m^2R_0^2=-3$ in the 2+1 dimensional
de Sitter worldsheet also corresponds to a massless field in 4+1
dimensional de Sitter space, considered in the previous subection. For
that reason, the mode solutions are very similar (except for an
overall power of the scale factor).

Note that the imaginary part of $W$, contributing to $|\Psi|^2$,
decays as $\tilde a^{-2}$. This is not surprising, since the normal
displacement is a tachyonic field whose amplitude grows with the scale
factor $\phi_{LM}(\tilde\eta) \propto \tilde a(\tilde\eta)$ at late
times.

To make contact with the boundary theory, we would like to consider
the relative co-moving displacement on the flat slices of constant
scale factor time. This is given by
\begin{equation}
\delta \equiv {\delta r\over r_w} = {1\over \gamma}{T^{-1/2} 
\phi \over \tilde a(t)}.\label{delta}
\end{equation}
The second factor in the right hand side is the proper normal
displacement (in the reference frame where the worldsheet is at rest)
divided by the physical radius of the bubble.  The relativistic
$\gamma$ factor accounts for the Lorentz contraction of the
displacement in the reference frame associated to the constant scale
factor hypersurfaces, and can be calculated from (\ref{normals}) as
\begin{equation}
\gamma = n^{\mu}\tilde n_{\mu} = R_0^{-1} r_w(t) e^{H t_0} 
\to (HR_0)^{-1}.\label{gamma}
\end{equation}
Here, $\tilde n_{\mu} = a(t) \delta_{\mu r}$ is the normal to a
co-moving sphere, and in the last step we have taken the late time
limit.

Note that $\gamma$ will be different on both sides of the wall if the
vacuum energy is different on both sides.  Geometrically, this means
that the surfaces of constant scale factor time on both sides of the
wall meet at an angle, and the corresponding frames are not at rest
with respect to each other.\footnote{Here we are neglecting the
gravitational field of the domain wall, so this kinematic effect is
unrelated to the jump in the extrinsic curvature as we go accross the
worldsheet, which would only be appreciable if the inequalities
(\ref{instgrav}) were violated.}  Although the unperturbed geometry of
the surfaces $\Sigma_t$ is flat, the embedding of these surfaces in
the bulk is non-trivial. Consequently, the relative perturbations
measured by co-moving observers on both sides are related by
\begin{equation}
H_{i}^{-1} \delta_{i} = H_{e}^{-1} \delta_{e},
\end{equation}
where the indices $i$ and $e$ refer to the interior and the
exterior. This means that in the presence of bubble fluctuations, we
cannot represent the future boundary as a smooth flat surface, because
the interiors of the perturbed bubbles do not fit nicely within the
holes they carve in the parent vacuum (the relative size of the
wiggles differs on both sides).  On the other hand, the bulk geometry
is smooth, and in principle we could choose to foliate the perturbed
bubble smoothly.  In this case, the interior of the bubble will match
the exterior, even as we approach the future boundary, but the metric
on the surfaces $\Sigma'$ of the new foliation would not be flat:
there would be some metric perturbations. In this sense, the situation
would not be so different from the case considered in the previous
subsection, where we allowed for metric perturbations on a de Sitter
background.

For the remainder of this Section, we shall avoid introducing curved
foliations by restricting attention to the case where the jump in
energy density accross the wall is sufficiently small,
\begin{equation}
{\Delta H \over H} \ll 1.
\end{equation}
In this limit, we can think of the bubble as propagating on a fixed
background de Sitter space characterized by a single parameter $H$, in
which case $\delta_e \approx \delta_i$. Using (\ref{delta}) and
(\ref{gamma}), Eq. (\ref{g3}) can be written as
\begin{equation}
W[\bar \delta] = 
{TR_0\over H^2} \sum_{LM}{\Delta (\Delta+2)\over 16}  
\left[\ln \left({R_0^2\over 4\tilde a^2}\right)+2\left(\psi(L)+
{1\over L}\right)+ i\pi+2\gamma\right]
\left|\delta_{LM}\right|^2 + ... \label{ww}
\end{equation}
where we have dropped the analytic divergent terms.

The expression vanishes both for $L=0$ ($\Delta=0$) and $L=1$
($\Delta=-2$).  This is in agreement with conformal invariance, which
requires that the effective action is independent of rescalings of the
bubble size and linearized translations (corresponding to $L=0$ and
$L=1$ respectively).

As mentioned in the previous subsection, for a CFT propagating in a curved background,
the coefficient of the logarithmic divergence is equal to the integrated trace anomaly 
coefficient $a_{d/2}$. When $d$ is odd and the manifold has no boundaries, $a_{d/2}$ vanishes
because we cannot build curvature invariants of odd
dimension. However, if the manifold has surfaces of co-dimension 1
where the CFT fields satisfy boundary conditions, then there is a
contribution to $a_{d/2}$ from geometric invariants constructed out of
the intrinsic and extrinsic curvatures, $\hat R_{abcd}$ and $K_{ab}$,
of the $p$-dimensional boundary (where $p=d-1$), as well as contractions
of the $d$-dimensional curvature $R_{ik}$ with the normal $n_i$ to the
boundary surface. If the boundary conditions are Weyl invariant, then
the corresponding coefficient $a_{d/2}$ is also Weyl invariant.

Let us now argue that the locus of the bubble wall at the future boundary 
plays the role of such a boundary surface.
The case of our interest is $p=2$, where the invariants
have to be of dimension 2, and so they must be linear in
$R_{ij}n^in^j$ or $\hat R$, or quadratic in the extrinsic curvature
$K_{ij}$. If $g_{ij}$ is flat, as we are assuming in this subsection
for the metric at the future boundary, then $R_{ij}=0$, and so the
integrated trace anomaly must be of the form
\begin{equation}
a_{3/2} = \int d\Sigma_2\left[d_1 \left(K_{ab}K^{ab}-{1\over
2}K^2\right) + d_2 \hat R \right].
\end{equation}
Here, $d\Sigma_2$ is the area element on the boundary surface and
$d_1$ and $d_2$ are constants.  It can be shown that the term
accompanying $d_1$ is Weyl invariant, while the term
accompanying $d_2$ is topological. It is easy to check that in
linearized theory around a spherical bubble wall, we have
\begin{equation}
\int d\Sigma_2 \left(K_{ab}K^{ab}-{1\over 2}K^2\right) \propto
\int d\Omega \ \delta \Delta (\Delta+2) \delta,
\end{equation}
where $\delta\equiv \delta r/r_w$ is the relative radial displacement at
the future boundary, and $\Omega$ is the solid angle. Note that the
coefficient of the logarithmic divergence in (\ref{ww}) has precisely
this form.  On the other hand, since the term proportional to $d_2$ is
topological, it does not depend on the perturbation $\delta$.

To conclude, the logarithmic divergence in (\ref{ww}) takes the form
expected in a CFT at the future boundary, where the domain walls play the role of surfaces where
the CFT fields have to satisfy boundary conditions.

The idea that domain walls act as boundary surfaces seems very natural
in the present context, since the number of field degrees of freedom
$c\sim H^{-2}$ is different on both sides of the wall.  The
coefficient $d_1$ can be read from (\ref{ww}),
\begin{equation}
d_1 \sim {TR_0\over H^2}.
\label{d1}
\end{equation}
Unless there are drastic cancellations, $d_1$ should be roughly equal
to the number of fields satisfying boundary conditions on the wall.
It follows from (\ref{instgrav}) that $d_1\ll H^{-2}$, indicating that
most of the fields pass freely through the bubble wall.

On the other hand, the number of fields satisfying nontrivial boundary
conditions on the wall should at least be equal to the difference
$\Delta c$ in the number of fields on the two sides of the wall. In
the absence of dramatic cancellations, this leads to 
\beq 
d_1\gtrsim
\Delta c \sim \Delta H/H^3 ,
\label{d1c}
\eeq
Substituting (\ref{d1}) into (\ref{d1c}), we obtain
\beq
TR_0 \gtrsim \Delta H/H.
\eeq
Using (\ref{radius}), the above inequality can only be satisfied provided 
that
\begin{equation}
\Delta\rho_V \lesssim H T. \label{bps}
\end{equation}
For such values of the parameters, the intrinsic curvature radius (or inverse proper acceleration) of
the bubble walls is comparable to the inverse Hubble radius of the parent vacuum
\begin{equation}
R_0 \sim H^{-1}. \label{rh}
\end{equation}

The necessity of the condition (\ref{bps}), and its possible
implications, are at present unclear to us.  It could be that there
are unexpected cancellations in the boundary theory which allow
$d_1\ll \Delta c$.
\footnote{Supersymmetry will not necessarily help enforcing the
  cancellations which would make $d_1$ much smaller than the number of
  fields satisfying boundary conditions at the wall.  For instance, in
  N=4 SYM in $d=4$ we have supersymmetry, but the coefficient of the
  trace anomaly is comparable to the number of fields.}  On the other
hand, it should also be noted that when (\ref{bps}) is violated, the
proper acceleration $R_0^{-1}$ of the bubble walls is much larger than
$H$, all the way to the future boundary. It could be that the
correspondence between bulk and boundary theory in this ``high
energy'' regime is not as straightforward as it seems to be for low
energy bubbles (with $R_0^{-1} \sim H$).  The investigation of this
issue is left for future research.

\section{Conclusions and discussion}

We have explored the conjecture, made in Ref.~\cite{GV08}, that the
inflationary multiverse has a dual holographic description at its
future boundary, in the form of a lower dimensional theory which is
conformally invariant in the UV.  The duality is expressed by the
relation
\beq
\Psi = e^{iW} ,
\label{duality}
\eeq
where $\Psi$ is the wave function of the multiverse with arguments on
a spacelike hypersurface $\Sigma$ and $W$ is the effective action of
the boundary theory on $\Sigma$.  Here, we have argued that the
boundary theory is defined on a ``fish net'' with characteristic
spacing set by the local horizon size, and that the number of degrees
of freedom in this theory is comparable to that in the bulk theory.  A
continuum description is obtained by imposing a Wilsonian cutoff, in
the limit where the cutoff length $\xi$ is large compared to all fish
net spacings.

To study the UV limit of the boundary theory, we foliate the bulk
spacetime (excluding terminal bubble interiors) with surfaces of
constant scale factor time $t$, starting with some initial surface
$\Sigma_0$.  (Geodesic crossings that may occur in structure formation
regions on sub-horizon scales do not interfere with this construction,
since the foliation surfaces are smoothed out with a super-horizon
cutoff $\xi$.)  As we go to larger values of $t$, the comoving fish
net scale decreases as $e^{-t}$, so we can choose the cutoff
$\xi(t)\propto e^{-t}$.  Thus, renormalization group flow in the
boundary theory corresponds to scale factor time evolution in the
bulk, with the UV limit $\xi\to 0$ on the boundary corresponding to
the IR limit $t\to \infty$ in the bulk.  In this limit, the foliation
surfaces approach the eternal set ${\cal E}$ at the future infinity.
Different choices of the initial surface $\Sigma_0$ are related by
Weyl rescalings in the boundary theory.  They should be physically
equivalent in the UV if the theory is indeed conformally invariant in
that limit.

To find evidence for (or against) the conformal invariance of the
boundary theory, we have studied a simple model in which the
inflating bubble interiors are pure de Sitter, so the inflating part
of spacetime consists of de Sitter regions separated by thin bubble
walls.  In this case the foliation surfaces can be chosen to be flat,
and we found that the bubble distribution on these surfaces is
approximately invariant under the Euclidean conformal group, with the
invariance becoming exact in the limit $t\to\infty$.

We have also studied the effect of linearized perturbations about the
model of nested dS bubbles.  Using the duality relation
(\ref{duality}), we have calculated the effective action $W$ for the
case of linearized tensor modes in de Sitter space and for
fluctuations of the bubble walls in the limit in which the gravity of
the wall is unimportant.  In both cases, the resulting functional form
of $W$ is consistent with that expected in a conformal field
theory. We interpret the locus of bubble walls at the future
  boundary as defining the surfaces where CFT fields must satisfy boundary
  conditions.  The form of the logarithmically divergent terms in $W$
  is in agreement with this interpretation.  Altogether, our results
support the conjecture that the boundary theory is conformally
invariant in the UV.

A puzzle arises for the case of bubble walls whose proper acceleration
$R_0^{-1}$ is much larger than the Hubble rate $H$ of the parent
vacuum. In this ``high energy'' regime, the numerical coefficient
$d_1$ in front of the logarithmic divergence in $W$ is much smaller
than the number of CFT fields which must satisfy boundary conditions
at the wall. This result is somewhat counterintuitive, and its
implications are left for further research.

Another important issue that has not been addressed in this paper is the
treatment of terminal bubbles.  We simply adopted the proposal of
Ref.~\cite{GV08} that the interiors of such bubbles should be excised
on the future boundary, with the expectation that their dynamics is
holographically described by $2d$ conformal field theories living
on the bubble boundaries.  Freivogel {\it et.~al.} \cite{FSY}
provided some evidence for this in the case of Minkowski bubbles.  For
an AdS bubble, most of the volume in the interior is near the bubble
wall, and it is natural to expect that a holographic description
should apply.  But as of now, this conjecture is not supported by any
quantitative evidence.

An objection that has often been raised against a holographic
description of de Sitter space is the anomalous behavior of massive
fields in the boundary theory.  Strominger \cite{Strominger1} has
studied the asymptotic behavior of the two-point function for a
massive scalar field at future infinity and concluded that the field
acquires a complex conformal weight if its mass is $m>3H/2$.  We note,
however, that there may be a good physical reason for this behavior.

Fields with $m>3H/2$ describe massive particles whose density is
diluted as $n\propto a^{-3}$, so the comoving number of particles is
conserved in the absence of interactions.  The particles, however, are
unstable and will decay in dS space, even if they are absolutely
stable in the flat space limit: the effective Gibbons-Hawking
temperature of the dS space allows decays in which the energy of the
decay products is higher than that of the initial particle.
Such particles cannot propagate any information to future infinity, so
there is little to be learned from the asymptotic properties of their
two-point function.  On the other hand, the comoving number of
particles with $m<3H/2$ grows with the scale factor time as
\beq
N \propto e^{\beta t}
\eeq
with
\beq
\beta = 3\left[1-\left({2m\over{3H}}\right)^2\right]^{1/2} .
\eeq
If their decay rate is not too high, such particles have a chance of
producing an imprint at future infinity.

As we mentioned in the Introduction, one of the main motivations for
studying the holographic description of the multiverse is its
potential relevance for the measure problem.  We have argued that a
Wilsonian UV cutoff in the boundary theory corresponds to a scale
factor cutoff on super-horizon scales in the bulk.  
The present work has also revealed some
subtleties in establishing the duality, particularly
in the case of bubbles which accelerate faster than the Hubble 
rate of the parent vacuum. Investigation of these issues 
may lead to a more detailed understanding of the boundary theory and its 
holographic relation to the bulk.

\section*{Acknowledgements}

We would like to thank Raphael Bousso, Stanley Deser, Roberto Emparan, Tomeu Fiol, Daniel Freedman, 
Ben Freivogel, Matthew Kleban, Igor Klebanov, Jose Ignacio Latorre, Juan Maldacena, Oriol Pujolas, 
Stephen Shenker and Leonard Susskind for useful discussions.
This work was supported in part by the Fundamental Questions Institute
(JG and AV), by grants FPA2007-66665C02-02 and DURSI 2005-SGR-00082
(JG), and by the National Science Foundation grant 0353314 (AV).

\section*{Appendix: Special Conformal Transformations and their 
relation to boosts.} 

In the flat chart of a dS space of unit radius, with metric given by
\begin{equation}
ds^2 = {\eta^{-2}}(-d\eta^2 + d{\bf x}^2),  \quad (-\infty < \eta < 0),
\end{equation} 
let us onsider the coordinate transformation
\begin{equation}
{x'^\mu \over x'^2} = {x^\mu \over x^2} - b^{\mu}, \label{isom}
\end{equation} 
where $x^{\mu} =(\eta,{\bf x})$ and $ b^{\mu}=(0, {\bf b})$.
Squares such as $x^2$ and $x'^2$, as well as the scalar products below, are constructed with
the Minkowski metric, $x\cdot y = -x^0y^0 +{\bf x}\cdot {\bf y}$.
We first note that since $b^0= 0$, we have
\begin{equation}
{x^2\over x'^2} = {\eta\over \eta'}. \label{rat1}
\end{equation}
Differentiating both sides of (\ref{isom}) and squaring them, we find 
${x'}^{-4} dx'^2= {x}^{-4} dx^2$. It follows that (\ref{isom}) is an isometry of dS
\begin{equation}
\eta'^{-2} (-d\eta'^2 + d{\bf x'}^2) = \eta^{-2} (-d\eta^2 + d{\bf x}^2).
\end{equation}
At the future boundary $\eta=\eta'=0$, the coordinate transformation
reduces to the Special Conformal Transformation (SCT):
\begin{equation}
{x'^i\over {\bf x'^2}} = {x^i\over {\bf x^2}} - {b^i}.
\label{sct}
\end{equation}

Since (\ref{isom}) is an isometry of dS, it must correspond to boosts and rotations in
the embedding Minkowski space with coordinates $(T,Y,\bf X)$, where de Sitter space is
given by the hypersurface
${\bf X}^2+Y^2-T^2=1$. Let us see this in
explicit form. Introducing the null coordinates $U=T-Y$ and $V=T+Y$, the embedding coordinates are given by
\begin{eqnarray}
U &=& -x^2/\eta,\\
V &=& -1/\eta,\\
{\bf X} &=& -{\bf x}/\eta.
\end{eqnarray}
Using (\ref{rat1}) we have
\begin{equation}
{V' \over V} = {x^2\over x'^2} = {1-2 b \cdot x +  b^2  x^2}.
\end{equation}
where in the last step we have used the square of Eq. (\ref{isom}). After some simple algebra,
we find
\begin{eqnarray}
U'&=&U,\\
V'&=& V - 2 {\bf b} \cdot {\bf X} + b^2 U,\\
{\bf X'}&=& {\bf X} - U{\bf b} .
\end{eqnarray}
Hence, in the embedding space, the transformation can be seen as a boost and a rotation, which mixes 
the spatial coordinates ${\bf X}$ with $Y$ and $T$.


\begin{thebibliography}{99}


\bibitem{Guth07} 
A.~H.~Guth, Phys. Rept. {\bf 333}, 555 (2000);
  ``Eternal inflation and its implications,''
J.\ Phys.\ A {\bf 40}, 6811 (2007) [arXiv:hep-th/0702178];

\bibitem{Winitzki08}
S.~Winitzki,
  ``Predictions in eternal inflation,''
Lect.\ Notes Phys.\ {\bf 738}, 157 (2008) [arXiv:gr-qc/0612164];

\bibitem{youngness2}
R.~Bousso, B.~Freivogel and I.~S.~Yang,
  ``Boltzmann babies in the proper time measure,''
arXiv:0712.3324 [hep-th].

\bibitem{LVW}
  A.~Linde, V.~Vanchurin and S.~Winitzki,
  ``Stationary Measure in the Multiverse,''
  JCAP {\bf 0901}, 031 (2009)
  [arXiv:0812.0005 [hep-th]].

\bibitem{DGSV}
  A.~De Simone, A.~H.~Guth, M.~P.~Salem and A.~Vilenkin,
  ``Predicting the cosmological constant with the scale-factor cutoff
  measure,''
  arXiv:0805.2173 [hep-th].

\bibitem{DeSimone:2008if}
  A.~De Simone, A.~H.~Guth, A.~Linde, M.~Noorbala, M.~P.~Salem and A.~Vilenkin,
  ``Boltzmann brains and the scale-factor cutoff measure of the multiverse,''
  arXiv:0808.3778 [hep-th].



\bibitem{FSY}
  B.~Freivogel, Y.~Sekino, L.~Susskind and C.~P.~Yeh,
  ``A holographic framework for eternal inflation,''
  Phys.\ Rev.\  D {\bf 74}, 086003 (2006)
  [arXiv:hep-th/0606204].

\bibitem{hat2}
  L.~Susskind,
  ``The Census Taker's Hat,''
  arXiv:0710.1129 [hep-th].

\bibitem{Bousso06}
  R.~Bousso,
  ``Holographic probabilities in eternal inflation,''
  Phys.\ Rev.\ Lett.\  {\bf 97}, 191302 (2006)
  [arXiv:hep-th/0605263]

\bibitem{GV08}
  J.~Garriga and A.~Vilenkin,
  ``Holographic Multiverse,''
  JCAP {\bf 0901}, 021 (2009)
  [arXiv:0809.4257 [hep-th]].

\bibitem{Bousso09}
  R.~Bousso,
  ``Complementarity in the Multiverse,''
  arXiv:0901.4806 [hep-th].

\bibitem{FK09}
  B.~Freivogel and M.~Kleban,
  ``A Conformal Field Theory for Eternal Inflation,''
  arXiv:0903.2048 [hep-th].



\bibitem{Maldacena98}
  J.~M.~Maldacena,
  ``The large N limit of superconformal field theories and supergravity,''
  Adv.\ Theor.\ Math.\ Phys.\  {\bf 2}, 231 (1998)
  [Int.\ J.\ Theor.\ Phys.\  {\bf 38}, 1113 (1999)]
  [arXiv:hep-th/9711200].

\bibitem{Gubser:98}
S.~S.~Gubser, I.~R.~Klebanov and A.~M.~Polyakov,
  ``Gauge theory correlators from non-critical string theory,''
  Phys.\ Lett.\  B {\bf 428}, 105 (1998)
  [arXiv:hep-th/9802109].

\bibitem{Witten98}
  E.~Witten,
  ``Anti-de Sitter space and holography,''
  Adv.\ Theor.\ Math.\ Phys.\  {\bf 2}, 253 (1998)
  [arXiv:hep-th/9802150].

\bibitem{recycling}  
J.~Garriga and A.~Vilenkin,
  ``Recycling universe,''
  Phys.\ Rev.\  D {\bf 57}, 2230 (1998)
  [arXiv:astro-ph/9707292].

\bibitem{GSVW}
  J.~Garriga, D.~Schwartz-Perlov, A.~Vilenkin and S.~Winitzki,
  ``Probabilities in the inflationary multiverse,''
  JCAP {\bf 0601}, 017 (2006)
  [arXiv:hep-th/0509184].

\bibitem{GGV}
J.~Garriga, A.~H.~Guth and A.~Vilenkin,
  ``Eternal inflation, bubble collisions, and the persistence of memory,''
  Phys.\ Rev.\  D {\bf 76}, 123512 (2007)
  [arXiv:hep-th/0612242].

\bibitem{discord}
  A.~Vilenkin,
  ``The wave function discord,''
  Phys.\ Rev.\  D {\bf 58}, 067301 (1998)
  [arXiv:gr-qc/9804051].

\bibitem{HawkingEllis}
S.W. Hawking and G.F.R. Ellis, ``The large scale Structure of
Spacetime", Cambridge (1973), Sect. 6.8.

\bibitem{SW98}
  L.~Susskind and E.~Witten,
  ``The holographic bound in anti-de Sitter space,''
  arXiv:hep-th/9805114.

\bibitem{Strominger1}
A.~Strominger,
  ``The dS/CFT correspondence,''
  JHEP {\bf 0110}, 034 (2001)
  [arXiv:hep-th/0106113];

\bibitem{Strominger2}
A.~Strominger,
  ``Inflation and the dS/CFT correspondence,''
  JHEP {\bf 0111}, 049 (2001)
  [arXiv:hep-th/0110087].

\bibitem{BVV}
J.~de Boer, E.~P.~Verlinde and H.~L.~Verlinde,
  ``On the holographic renormalization group,''
  JHEP {\bf 0008}, 003 (2000)
  [arXiv:hep-th/9912012].

\bibitem{VS}
  J.~P.~van der Schaar,
  ``Inflationary perturbations from deformed CFT,''
  JHEP {\bf 0401}, 070 (2004)
  [arXiv:hep-th/0307271].

\bibitem{LMN}

F.~Larsen and R.~McNees,
  ``Inflation and de Sitter holography,''
  JHEP {\bf 0307}, 051 (2003)
  [arXiv:hep-th/0307026].




\bibitem{Sasaki}
See e.g. H.~C.~Lee, M.~Sasaki, E.~D.~Stewart, T.~Tanaka and S.~Yokoyama,
  ``A new delta N formalism for multi-component inflation,''
  JCAP {\bf 0510}, 004 (2005)
  [arXiv:astro-ph/0506262].


 
\bibitem{Maldacena02}
  J.~M.~Maldacena,
  ``Non-Gaussian features of primordial fluctuations in single field
  JHEP {\bf 0305}, 013 (2003)
  [arXiv:astro-ph/0210603].

\bibitem{Tomboulis:77}
  E.~Tomboulis,
  ``1/N Expansion And Renormalization In Quantum Gravity,''
  Phys.\ Lett.\  B {\bf 70}, 361 (1977).

\bibitem{BD} See e.g. N.D. Birrell and P.C.W. Davies, ``Quantum fields in curved space'', 
Cambridge (1982), and references therein.

\bibitem{Berezin:87}
  V.~A.~Berezin, V.~A.~Kuzmin and I.~I.~Tkachev,
  Phys.\ Rev.\  D {\bf 36}, 2919 (1987).

\bibitem{rates}
  J.~Garriga,
  ``Nucleation rates in flat and curved space,''
  Phys.\ Rev.\  D {\bf 49}, 6327 (1994)
  [arXiv:hep-ph/9308280].

\bibitem{GV91}
  J.~Garriga and A.~Vilenkin,
  ``Perturbations on domain walls and strings: A Covariant theory,''
  Phys.\ Rev.\  D {\bf 44}, 1007 (1991);
J.~Garriga and A.~Vilenkin,
  ``Quantum fluctuations on domain walls, strings and vacuum bubbles,''
  Phys.\ Rev.\  D {\bf 45}, 3469 (1992).




\end{thebibliography}
\end{document}